\documentclass[10pt]{iopart}
\usepackage{iopams}
\usepackage[usenames,dvipsnames]{xcolor} 
\usepackage[normalem]{ulem}
\usepackage[labelfont=bf,font=small]{caption}
\usepackage{graphicx}
\usepackage{iopams}
\usepackage{epstopdf}

\begin{document}

\note{Linear Degeneracy of the First-Order Generalized-Harmonic
  Einstein System}

\author{ Li-Wei Ji$^{1}$, Lee Lindblom$^{2}$, Zhoujian Cao$^{3}$}
\address{${}^{1}$Department of Astronomy, Beijing Normal University, Beijing,
  100875, China}
\address{${}^{2}$Center for Astrophysics and Space Sciences,
  University of California at San Diego, La Jolla, CA 92093, USA}
\address{${}^{3}$ Institute of Applied Mathematics and the
  Laboratory of Scientific and Engineering Computing,
  Academy of Mathematics and Systems Science,
  Chinese Academy of Sciences, Beijing 100190, China}

\begin{abstract}
  The purpose of this note is to clarify the conditions under which
  the first-order generalize-harmonic representation of the vacuum Einstein
  evolution system is linearly degenerate.
\end{abstract}

\submitto{\CQG}

\maketitle

The formation of coordinate shocks is one of the important problems
that must be overcome by any representation of Einstein's equation
that is to be used successfully in numerical relativity.  Poor
dynamical gauge conditions can and will lead to the formation of
shocks (and consequently coordinate singularities) from the evolution
of smooth initial data~\cite{Alcubierre1997}.  Linear degeneracy is a
mathematical condition that prevents the formation of shocks in a
large class of hyperbolic evolution
systems~\cite{lax1973hyperbolic,liu1979,Li1994}.  The purpose of this
note is to clarify the conditions under which the first-order
generalized-harmonic representation of the vacuum Einstein
system~\cite{Lindblom:2005qh} is linearly degenerate.  The original
paper on this system claimed, without presenting a proof, that the
system was linearly degenerate if a certain constant satisfied the
condition $\gamma_1=-1$~\cite{Lindblom:2005qh}.  Here we demonstrate
that this claim is correct.  While the proof is fairly
straightforward, some readers of the original paper have questioned
whether that condition is correct~\cite{cao2018application}.
Consequently it seems appropriate to provide a more complete
description of the derivation that demonstrates this fact.

The first-order generalized-harmonic representation of Einstein's
vacuum equation~\cite{Lindblom:2005qh} can be written abstractly as a
quasi-linear system,
\begin{equation}
  \partial_tu^\alpha+A^{k\alpha}{}_\beta\partial_ku^\beta=F^\alpha,
  \label{e:quasilineareq}
\end{equation}
where $u^\alpha=\{\psi_{ab},\Pi_{ab},\Phi_{iab}\}$ is the collection
of dynamical fields: $\psi_{ab}$ the spacetime metric, and its time
and space derivatives $\Pi_{ab}$ and $\Phi_{iab}$.  The quantities
$A^{k\alpha}{}_\beta$ and $F^\alpha$ depend on $u^\alpha$ but not its
derivatives.  We use the notation $s^\alpha \,\partial_{u^\alpha}$ for
vectors, and $t_\alpha\,du^\alpha$ for co-vectors on the space of
dynamical fields.  The principal parts of the first-order
generalized-harmonic vacuum Einstein system, $\partial_tu^\alpha +
A^{k\alpha}{}_\beta\partial_ku^\beta\simeq 0$, are given explicitly by
\begin{eqnarray}
\partial_t\psi_{ab}
-(1+\gamma_1)N^k\partial_k\psi_{ab} &\simeq& 0,
\label{e:NewPsiDot}\\
\partial_t\Pi_{ab} - N^k\partial_k\Pi_{ab}
+ N g^{ki}\partial_k\Phi_{iab} -
\gamma_1\gamma_2 N^k\partial_k\psi_{ab}&\simeq &0,
\label{e:NewPiDot}\\
\partial_t\Phi_{iab}-N^k\partial_k\Phi_{iab}
+N\partial_i\Pi_{ab}-\gamma_2 N \partial_i\psi_{ab}&\simeq&0,
\label{e:NewPhiDot}
\end{eqnarray}
where $N$ and $N^k$ are the lapse and shift associated with the
standard $3+1$ representation of the metric $\psi_{ab}$, and where
$\gamma_1$ and $\gamma_2$ are constants.\footnote{The constants
  $\gamma_1$ and $\gamma_2$ multiplied by certain constraints of the
  vacuum Einstein system were added to the equations in
  Ref.~\cite{Lindblom:2005qh}.  The resulting system is symmetric
  hyperbolic for any values of these constants.  As shown here, the
  constant $\gamma_1$ effects the linear degeneracy of the system. The
  constant $\gamma_2$ effects the growth of small constraint
  violations, and must be positive, $\gamma_2>0$, for numerical
  stability.} The characteristic matrix $n_kA^{k\alpha}{}_\beta$ for
this system can be written as
\begin{eqnarray}
&&  n_kA^{k\alpha}{}_\beta\, \partial_{u^\alpha}\otimes d u^\beta =
  -(1+\gamma_1)n_kN^k\, \partial_{\,\psi_{ab}}\!\otimes d\psi_{ab}
  \nonumber\\
  &&\,\,\,
  -n_kN^k\,\partial_{\,\Pi_{ab}}\!\otimes d\Pi_{ab}
  +N n^k\,\partial_{\,\Pi_{ab}}\!\otimes d\Phi_{kab}
  -\gamma_1\gamma_2 n_kN^k\,\partial_{\,\Pi_{ab}}\!\otimes d\psi_{ab}
  \nonumber\\
  &&\,\,\,
  -n_k N^k\,\partial_{\,\Psi_{iab}}\!\otimes d\Psi_{iab}
  +N n_k\,\partial_{\,\Psi_{kab}}\!\otimes d\Pi_{ab}
  -\gamma_2N n_k\,\partial_{\,\Psi_{kab}}\!\otimes d\psi_{ab},\nonumber\\
\end{eqnarray}
for waves propagating through a surface (chosen arbitrarily) with
spacelike unit normal co-vector $n_k$. Summation over repeated indices, e.g.,
$\scriptstyle{k}$, $\scriptstyle{a}$, and $\scriptstyle{b}$, is
implied.  Linear degeneracy is a condition on the eigenvalues and
eigenvectors of this characteristic matrix.

The left and right eigenvectors, $\ell^{\,\hat\alpha}$ and
$r_{\hat\alpha}$ respectively, of the characteristic matrix
$n_kA^{k\alpha}{}_\beta$ are defined by:
\begin{eqnarray}
  \ell^{\,\hat\alpha}{}_\alpha n_kA^{\,k\alpha}{}_\beta &=& v^{(\hat\alpha)}
  \ell^{\hat\alpha}{}_\beta,\\
   n_kA^{\,k\alpha}{}_\beta r_{\hat\alpha}{}^\beta&=& v_{(\hat\alpha)}
  r_{\hat\alpha}{}^\alpha.
\end{eqnarray}
The left eigenvectors $\ell^{\,\hat\alpha}{}_\beta du^\beta$ of the
first-order generalized-harmonic vacuum Einstein system are given by
\begin{eqnarray}
  \ell^{\,\hat 0}_{ab\,}{}_\beta\, du^\beta &=& d\psi_{ab},\\
  \ell^{\,\hat 1\pm}_{ab\,}{}_\beta\, du^\beta &=&
  d\Pi_{ab}\pm n^id\Phi_{iab}-\gamma_2\, d\psi_{ab},\\
  \ell^{\,\hat 2}_{iab\,}{}_\beta\, du^\beta &=&
  (\delta_i{}^j - n_i n^j) d\Phi_{jab},  
\end{eqnarray}
while the right eigenvectors $r_{\hat\alpha}{}^\beta \partial_{u^\beta}$ are given by
\begin{eqnarray}
  r_{\hat 0}^{ab\,}{}^\beta\,\partial_{u^\beta} &=&
  \partial_{\psi_{ab}}+\gamma_2\,\partial_{\Pi_{ab}},
  \\
  r_{\hat 1\pm}^{ab\,}{}^\beta \,\partial_{u^\beta} &=&
  \partial_{\Pi_{ab}}\pm n_i\,\partial_{\Phi_{iab}},
  \\
  r_{\hat 2}^{iab\,}{}^\beta\,\partial_{u^\beta} &=&
  (\delta^i{}_j-n^in_j)\,\partial_{\Phi_{jab}}.
\end{eqnarray}
The first-order vacuum Einstein system is symmetric hyperbolic, since
there exists a symmetric positive definite tensor $S_{\alpha\beta}$ on
the space of fields that satisfies the condition
$S_{\alpha\mu}A^{k\mu}{}_\beta\equiv
A^{k}{}_{\alpha\beta}=A^k{}_{\beta\alpha}$~\cite{Lindblom:2005qh}.
This implies that the left and right eigenvectors are related (up to
normalizations) by $\ell^{\hat\alpha}{}_\alpha =
S_{\alpha\beta}r_{\hat\alpha}{}^\beta$, and the associated eigenvalues
must be equal $v_{(\hat\alpha)}=v^{(\hat\alpha)}$.  These eigenvalues
for the vacuum Einstein system are given by
\begin{eqnarray}
  v^{(\hat 0)}&=v_{(\hat 0)}&=-(1+\gamma_1)n_kN^k, \label{e:eig0}\\
  v^{(\hat 1\pm)}&=v_{(\hat 1\pm)}&=\pm N - n_kN^k,\label{e:eig1}\\
  v^{(\hat 2)}&=v_{(\hat 2)}&=-n_kN^k.\label{e:eig2}
\end{eqnarray}

The quasi-linear hyperbolic evolution system,
Eq.~(\ref{e:quasilineareq}), is said to be linearly degenerate if all
the eigenvalues of the characteristic matrix are constant along the
corresponding right eigenvectors of that system, so that
\begin{equation}
  r_{\hat\alpha}{}^\alpha\, \frac{\partial v_{(\hat\alpha)}}{\partial u^\alpha} = 0,
  \label{e:lineardegen}
\end{equation}
for each $\scriptstyle{\hat\alpha}$~\cite{lax1973hyperbolic}.  The
eigenvalues of the Einstein system,
Eqs.~(\ref{e:eig0})--(\ref{e:eig2}), depend only on the lapse, $N$,
the shift, $N^k$, and the unit normal vector $n^k$.  These eigenvalues
therefore depend only on the metric, $\psi_{ab}$, and not on its
derivatives, $\Pi_{ab}$ or $\Phi_{iab}$.  Thus the derivatives of the
eigenvalues in the direction of the right eigenvectors are given by,
\begin{eqnarray}
  r^{ab\,}_{\hat 0}{}^\alpha\,\frac{\partial v_{(\hat 0)}}{\partial u^\alpha}
  &=&\frac{\partial v_{(\hat 0)}}{\partial \psi_{ab}} = -(1+\gamma_1)
  \frac{\partial(n_kN^k)}{\partial\psi_{ab}},\\
  r^{ab}_{\hat 1\pm}{}^\alpha\,\frac{\partial v_{(\hat 1\pm)}}{\partial u^\alpha}&=&0,\\
  r^{iab\,}_{\hat 2}{}^\alpha\,\frac{\partial v_{(\hat 2)}}{\partial u^\alpha}&=&0.
\end{eqnarray}
These derivatives vanish, and consequently the system is linearly
degenerate, if and only if $\gamma_1=-1$.

The original paper on the first-order generalized-harmonic vacuum
Einstein system did not explicitly give expressions for either the
left or the right eigenvectors~\cite{Lindblom:2005qh}.  The
characteristic fields, $\hat u^{\hat\alpha}\equiv
\ell^{\hat\alpha}{}_\beta u^\beta$, of this system were given,
however, and from those the left eigenvectors could easily be
inferred.  The reported confusion about the correct conditions for
linear degeneracy for this system may have arisen from an examination
of the quantities $\ell^{\hat\alpha}{}_\alpha\partial
v^{(\hat\alpha)}/\partial u^\alpha$, which do not vanish unless
$\gamma_1=-1$ and $\gamma_2=0$~\cite{cao2018application}.  These
quantities involving the left eigenvectors (which are co-vectors, not
true vectors) are non-covariant and are therefore meaningless from a
fundamental mathematical viewpoint.  In any case they are irrelevant
because the formal definition of linear degeneracy given by Lax in
Ref.~\cite{lax1973hyperbolic} specifies that the right eigenvectors
are to be used in Eq.~(\ref{e:lineardegen}), and this equation is covariant.

\ack L.L. thanks the Morningside Center for Mathematics, Academy of
Mathematics and Systems Science, Chinese Academy of Sciences, Beijing
100190, China for their hospitality during a visit in which a portion
of this research was performed.  This research was supported in part
by the National Science Foundation grants PHY-1604244 and DMS-1620366
to the University of California at San Diego.

\section*{References}
\bibliographystyle{iopart-num}
\bibliography{refs}

\end{document}